# ESAS: An Efficient Semantic and Authorized Search Scheme over Encrypted Outsourced Data

Xueyan Liu[1], Zhitao Guan[1], Xiaojiang Du[2], Liehuang Zhu[3], Zhengtao Yu[4], Yinglong Ma[1]

1. School of Control and Computer Engineering, North China Electric Power University, Beijing, China
2. Department, Zhitao Guan[1] of Computer and Information Sciences, Temple University, Philadelphia, PA, USA
3. School of Computer, Beijing Institute of Technology, Beijing, China
4. School of Information Engineering and Automation, Kunming University of Science and Technology, Kunming, Yunnan, China

*Abstract*— Nowadays, a large amount of user privacy-sensitive data is outsourced to the cloud server in ciphertext, which is provided by the data owners and can be accessed by authorized data users. When accessing data, the user should be assigned with the access permission according to his identities or attributes. In addition, the search capabilities in encrypted outsourced data is expected to be enhanced, i.e., the search results can better present user's intentions. To address the above issues, ESAS, an Efficient Semantic and Authorized Search scheme over encrypted outsourced data, is proposed. In ESAS, by integrating PRSCG (the privacy-preserving ranked search based on conceptual graph) and CP-ABE (ciphertext policy attribute-based encryption), semantic search with file-level fine-grained access authorization can be realized. In addition, search authorization can be done in an offline manner, which can improve search efficiency and reduce the response time. The security analysis indicate that the proposed ESAS meets security requirement.

## I. INTRODUCTION

Today, cloud computing has become a mature computing model with a wide range of applications. And as an extension of the emerging concept of cloud computing, cloud storage has partially addressed the challenges of massive data storage. Enterprise and individual users purchase cloud services from cloud service providers and utilize cloud services for data management and storage. However, cloud service providers, which are not completely trusted by users, may suffer from internal or external attacks. People may be more concerned about data privacy issues, which will become a hindrance to the development of cloud computing. Since encryption is an effective way to protect data, ciphertext search technology has become a hot topic in the current research field.

Recently, searchable encryption techniques have been used to solve the problem of the utilization of encrypted cloud data. The related investigation has developed from traditional keyword-based search to smart semantic search. Traditional keyword-based search schemes can hardly satisfy people's search requirements in nowadays since people tend to semantic search which can better understand user's intention. The most advanced smart semantic search scheme over encrypted outsourced data PRSCG [1] employees Conceptual Graph (CG) [2] as its' semantic carrier, which is a perfect and mature manifestation of semantics , and realizes simplified quantitative calculation by linearization.

In a distributed cloud environment, a large amount of privacy sensitive data is outsourced to the cloud in ciphertext, which is provided by the data contributors and can be accessed by authorized data users. When accessing data, the user should be assigned with the access permission according to his identities or attributes. This makes file-level fine-grained authorization a problem. Existing solutions such as Sun's [3] and Li's [4] have some drawbacks. Sun's scheme cannot achieve ranking search and needs a large computational overhead. In Li's scheme, the search response time could be very long due to a low hit rate. That is to say, it takes a long time from the user issues a search query to obtain the desired results .This will significantly affect the user's search experience. And neither of them can implement semantic search.

To solve the above problems, we propose our efficient semantic authorized search schemes ESAS by integrating PRSCG [1] and CP-ABE [5] .The contributions of our work as follows:

*1)* By integrating PRSCG [1] and CP-ABE [5], we propose the ESAS scheme which uses Conceptual Graphs (CGs) as our semantic carrier to implement semantic search with file-level fine-grained access authorization.

*2)* Search authorization is done in an offline manner. Search authentication is preprocessed before the user's query, which improves search efficiency and reduces the response time.

*3)* The security analysis indicate that our system meets security requirements.

The rest of this paper is organized as follows. Section II introduces the related work. In section III, the problem formulation is given. In section IV, our scheme is described in detail. In section V, the security analysis is stated. In Section VI, the paper is concluded.

## II. RELATED WORK .

Recently, searchable encryption has been used to solve the problem of the utilization of encrypted cloud data. The related investigation has developed from traditional keyword-based search to semantic search. Key management [6-8] is also important for the related security schemes.

In [6], Song et al. proposed the symmetric ciphertext scheme for the first time. In [10], Cao et al. presented a novel multi-keyword search framework MRSE supporting ranked search. The scheme was proposed under two different attack models in which the relevance scores were calculated in two methods by leveraging KNN-SE algorithm [11] . In addition, the vectors are extended and virtual keywords are inserted to meet privacy requirements. Extended schemes [12], [13], [14] were proposed based on [10]. Fu et al. in [12] achieved personal search and semantic search with the help of online natural language processing tool WordNet. They constructed an interest model for each legal data user to store their personal intentions and extend the data user's query semantically to obtain more related results. Fu et al. in [13] proposed a scheme that supported fuzzy search and ranked search simultaneously. In [14], Xia et al. constructed the indexes using binary balanced trees, which improves the search efficiency and supports dynamic updates of documents.

At present, research on authorized keyword search has been carried out a lot. Yang et al. in [15] employed a user list that is server-enforced to assist completing the search in the practical application scenario to achieve search authorization. The user lists contain all complementary keys of the legal users. But the scheme based on SKC only supports one data contributor. Sun et al.[3] put forward a ciphertext search scheme in a multi-contributor multi-requestor scenario. It achieves fine-grained search authorization which is owner-enforced with efficient user and attribute revocation by leveraging CP-ABE [5] technique. The search results can be verified in terms of correctness, integrity and freshness. However, the scheme proposed in this paper only supports the exact search of multi-keyword, cannot realize semantic search and ranked search. Li et al. [4] proposed a scheme that achieves file-level fine-grained authorization and ranking search simultaneously which reduces computational overhead compared to [3]. In the distributed cloud environment, the attribute-based encryption (ABE) [16] is widely used in various networks because of its flexible access control and allowing users' data sharing at fine-grained level. Reference [17] defines the attribute's revocation in CP-ABE with semi-trusted servers. Several papers (e.g., [18-22]) have studied related security and cloud issues.

## III. PROBLEM FORMULATION

### A. System Model

The efficient semantic and authorized search scheme over encrypted cloud data ESAS is proposed by integrating the PRCSG [1] and CP-ABE [5]. The architecture of our scheme is shown in Fig. 1. There are six entities:

- **Data Owner (DO)** provides documents to be outsourced and assigns an access policy for each document.
- **Data User (DU)** generates a trapdoor to issue data retrieval request to AMS and CSP. The data owner also specifies the number of results expected to be returned
- **Key Generation Center (KGC)** is a trusted server. It takes a security parameter as input to generate all public and secret keys in the system.
- **Cloud Service Provider (CSP)** provides storage and computing services. CSP calculates top-k highest score results according to search requests and return them to data users.
- **Authority Management Server (AMS)** is a trusted server. It is responsible for search authorization for legitimate users.
- **Natural Language Processing Server (NLPS)** is a user-side server. It is responsible for semantic analysis of documents and secure index generation.

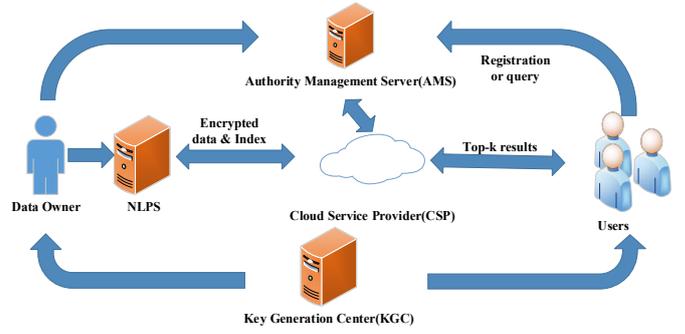

Fig.1. The architecture of the ESAS scheme

### B. Security Goals

In our ESAS scheme, we make the assumption that CSP is honest-but-curious, i.e., it is curious about users' privacy data and tries to deduce some useful information. Data users in our scheme may collude to access unauthorized documents. The security goals of our system are defined as follows.

*1) Confidentiality of indexes and documents.* Our scheme should protect the confidentiality of doucuments and indexes. Since CSP is semi-trusted in the system, it is curious about the privacy information of users. CSP should not deduce any useful information from the secure indexes and ciphertexts of documents.

*2) Unlinkability of Query.* Our scheme should ensure that the trapdoors genreaterd by queries are unlinkable. That's to say, even if the data user issues two identical queries , they can be transformed into two completely different trapdoors. CSP cannot deduce the relationship between trapdoors.

*3) Collusion resistance.* We should make sure that malicious users in the system can not combine their secret key components to get more privileges they would not have otherwise.

### C. Notations

In table I, we list the notations used in ESAS.

TABLE I.  NOTATIONS IN ESAS

| Acronym | Descriptions |
|---|---|
| $A_i$ | The i-th data owner. |
| $B_i$ | The i-th data user. |
| $D_{i,j}$ | The j-th document of $A_i$ |
| PP | Public Prameter |
| $SK_{A_i}$ | The secrect key of data owner $A_i$ |
| $SK_{B_i}$ | The secrect key of data owner $B_i$ |
| GCs | Conceptual Graphs |
| T | Access Tree |
| CK | The ciphertext component of k |
| Vc | Vectors corresponding to the source document |
| Td | Trapdoor |
| Ic | the secure index |

## D. Assumption

We make two assumptions in our EASA scheme. They are as follows.

Assumption 1: Cloud Service Providers are semi-trusted. That is, they will follow our proposed protocol in general, but try to find out as much secret information as possible.

Assumption 2: KGC and AMS are trusted by all entities in the system. They will never leak information about data and related keys.

In our scheme, we make an assumption that the CSP will never collude with the malicious users to help them obtain unauthorized access privileges.

## IV. ESAS

The ESAS scheme is described in detail in this section. Six phases are included in ESAS: Initialization, User Registration, Secure Index Construction, Search Authorization, Trapdoor Generation, Query.

### A. Initialization

KGC generates two multiplicative cyclic group $G_1$ and $G_2$ according to security parameter of the system. Then, it produces a generator $g$ of $G_1$, a bilinear map e: $G_1 \times G_1 \to G_2$, collusion-resistant hush functions H:$\{0,1\}^* \to G_1$, $H_1: G_1 \to \kappa$. KGC also chooses system's secret parameters $\alpha, \beta, a, b \in Z_p$ randomly. Let $X$ be $e(g,g)^\alpha$, the public parameters of the system is

$$PP = \{H, G_1, G_2, g, e, g^a, g^b, X\}$$

### B. User Registration

If data owner $A_i$ applies to join the system, KGC generates secret key $SK_{A_i} = \{S, M_1, M_2\}$ which includes one (n+1)-bit binary vector S and two (n+1) × (n+1) invertible matrices $M_1, M_2 \in R^{n+1} \times R^{n+1}$. When data user $B_i$ applies registration for the first time, KGC generates $B_i's$ secret key which is related to his attribute set $At_{B_i}$. Firstly, KGC selects a random $u \in Z_p$ and computes $S = g^{(\alpha+u)/\beta}$, $S_1 = g^u$ then it chooses $r_m \in Z_p$ for each attribute $at_m \in At_{B_i}$ and calculates $S_m = g^u \cdot H(at_m)^{r_m}$, $S_m' = g^{r_m}$ The secret key of data user $B_i$ $SK_{B_i} = \{S, S_1, \forall m \in At_{B_i} : (S_m, S_m')\}$. It is sent to the corresponding user via a secure channel.

### C. Security Index Construction

If data owner $A_i$ wants to outsource one of his document $D_{i,j}$ to the cloud, he chooses a random number $k_{i,j} \in G_1$ and encrypts $D_{i,j}$ to $[D_{i,j}]_{\kappa_{i,j}}$ using symmetric encryption technologies. The symmetric encryption key $\kappa_{i,j}$ is generated by $H_1(K_{i,j})$. Then, $A_i$ assigns an access policy for $D_{i,j}$. He selects two random numbers $s, r_0 \in Z_p$ and set $r_0$ as the root secret value to generate the access tree $T_{i,j}$ [23]. Hereafter, $A_i$ computes $C = KX^{r_0}$, $C_0 = g^s$, $C_1 = g^{\beta r_0}$, $C_2 = g^{s-r_0}$ and for each leaf node n in $T_{i,j}$, $C_n = g^{q_n(0)}$, $C_n' = (H(att(n))^{q_n(0)}$. Let N be the leaf node set of $T_{i,j}$. The ciphertext component of $\kappa_{i,j}$ is defined as

$$CK_{i,j} = \{T_{i,j}, C, C_0, C_1, C_2, \forall n \in N : (C_n, C_n')\}. \quad (1)$$

Finally, data owner $A_i$ sends $D_{i,j}$ to NLPS for semantic analysis and index generation. NLPS processes the documents in a few steps as follows.

**Step1** NLPS chooses a processing method according to the characteristics of each document content as described in [1]. One of the methods is to calculate the scores of all the sentences and select the sentence with the highest score as the theme of the document. Then, NLPS simplifies the single theme sentence using Tregex [1] and expresses it in the form of a Conceptual Graph [2]. This method is for documents with clear topics. In another method, NLPS generates a conceptual graph for every sentence in the document. Through the above two methods, each document is transformed into one or a set of CGs.

**Step2** Conceptual Graph (CG) is an excellent knowledge representation tool with better semantic expressiveness than simple keywords. To prediget the matching process of CGs, we represents the CGs in a linear form [1]. Starting from the vertices with the most edges, a CG is divided into several entities, each of which is a ternary group and can be represented as a dimension in the vector space model. For example, the sentence "Amy is going to London by train" can be expressed as triples: $[Go, Dest, London]$, $[Go, Agent, Amy]$, and $[Go, Inst, train]$.

**Step3** NLPS constructs vectors $V_{i,j}$ for each document $D_{i,j}$ with the help of linear CGs. In the first method of Step1, $V_{i,j}$ is a binary vector. And in the second method, each dimension of $V_{i,j}$ is the value of $TF \times IDF$. TF represents the frequency of an entity in the document, and IDF reflects the significance of the entity in the whole documents.

**Step4** NLPS encrypts the vectors acquired by CGs to generate secure indexes. Specifically, $Vc_{i,j}$ should be extended to n+1 dimensions and recorded as $Vc_{i,j}^*(n,:)$. The n+1 dimension of $Vc_{i,j}^*(n,:)$ is 1. Then $Vc_{i,j}^*(n,:)$ is splitted into two random vectors $(Vc_{i,j}', Vc_{i,j}'')$ follows the rules: if $S[t]$ ($t \in [1, n+1]$) equals to 0, $Vc_{i,j}'[t]$ and $Vc_{i,j}''[t]$ will be the same as $Vc_{i,j}^*[t]$; else $Vc_{i,j}'[t]$ and $Vc_{i,j}''[t]$ will be set as random numbers which satisfies $Vc_{i,j}'[t] + Vc_{i,j}''[t] = Vc_{i,j}^*[t]$. The secure index $Ic_{i,j} = \{M_1^T Vc_{i,j}', M_2^T Vc_{i,j}''\}$.

After the construction of secure indexes, $A_i$ sends $E_{i,j} = (ID_{i,j}, CK_{i,j}, Ic_{i,j}, [D_{i,j}]_{K_{i,j}})$ to CSP and $(ID_{i,j}, CK_{i,j})$ to AMS.

### D. Search Authorization

AMS automatically performs search authentication for each legitimate user in the system. When a data user $B_i$ joins the system and completes registration, AMS authenticates search rights for each document and generates a list $L_{B_i}$ which stores the identities of documents he can access. Firstly, AMS tries to match each document's access tree using $B_i$'s secret key. If there exists an attribute of $B_i$ matches leaf node n in the access tree, AMS computes

$$Dec(SK, CK, m, n) = \frac{e(S_m, C_n)}{e(S_m', C_n')} \qquad (1)$$
$$= e(g,g)^{u \cdot q_n(0)} = Y^{q_n(0)} M_n$$

Let Y be $e(g,g)^u$. Then, AMS combines all $M_n$ to calculate the secret value $M_r = e(g,g)^{ur_0}$ of the access tree's root node as described in [5]. If the following equation holds, AMS adds the ID of the corresponding document to the list of accessible documents $L_{B_i}$.

$$e(C_2, S_1) M_r = e(g^{s-r_0}, g^u) \cdot Y^{r_0} \qquad (2)$$
$$= e(g^s, g^u) = e(C_0, S_1)$$

When the data owner outsource a document to the cloud, AMS refreshes all data users' $L_{B_i}$.

### E. Trapdoor Generation

The process of trapdoor generation is similar to index construction. The data user obtains an n-dimensional vector $Q$ by processing the query statement. The vector $Q$ is extended to (n+1) dimensions and changed into $Q^* = (aQ, r')$ (a and $r'$ are random numbers and $r'$ cannot be 0). Then $Q^*$ is divided into two vectors $(Q', Q'')$ which follows: if $S[t]$ ($t \in [1, n+1]$) equals to 0, $Q'[t]$ and will be set as random numbers and $Q'[t] + Q''[t] = Q^*[t]$; else $Q'[t]$, $Q''[t]$ will be the same as $Q^*[t]$. The trapdoor can be constructed as $Td = \{M_1^{-1} Q', M_2^{-1} Q''\}$.

### F. Query

If the data user issues a search request, he sends the trapdoor $Td$ to CSP and his signature $Sig(Q, SK_{B_i})$ to AMS.

Upon receiving the user's search request, AMS authenticates user's identity by his signature. Then, AMS sends $L_{B_i}$ to CSP. CSP calculates the relevance scores of each document in the list with the query by

$$Ic_{i,j} \cdot Td = \{M_1^T Vc', M_2^T Vc''\} \cdot \{M_1^{-1} Q', M_2^{-1} Q''\} \qquad (3)$$
$$= a(Vc \cdot Q) + r'$$

By sorting all relevance scores, CSP picks up k results with the highest scores and sends corresponding $E_{i,j}$ to the data user. Upon receiving the set of $E_{i,j}$, the data user firstly recovers the value of $M_r$ for each $E_{i,j}$. Then, he acquires every document's decryption key by computing

$$K_{i,j} = \frac{C}{e(C_1, S)/Mn}$$
$$= \frac{K \cdot X^{r_0}}{e(g^{\beta r_0}, g^{(\alpha+u)/\beta})/Y^{r_0}} \qquad (4)$$

After recovering $K_{i,j}$ for each document, the data user calculates the corresponding symmetric key by calculating $H_1(K_{i,j})$.

## V. SECURITY ANALYSIS

In this section, we analyze the security of our scheme from the three aspects: confidentiality of indexes and documents, unlinkability of query and collusion resistance.

### 1) Confidentiality of indexes and documents.

Our scheme should protect the confidentiality of doucuments and indexes. Since CSP is semi-trusted in the system, it is curious about the privacy information of users. CSP should not deduce any useful information from the secure indexes and ciphertexts of documents. The confidentiality of indexes is based on the security of KNN-SE algorithm[11] and CP-ABE. For PRSCG, its security is based on MRSE[10]. That's to say, we can guarantee the privacy of indexes if KNN-SE can be proved to be secure. In this way, CSP cannot deduce any useful information from the secure indexes. Since the documents are encrypted by symmetric keys generated by a hash function, we base the confidentiality of documents on the security of symmetric encryption. The ciphertexts of documents can be decrypted only if the symmetric key is obtained. Because the symmetric key is protected by CP-ABE [5], only data users whose attributes satisfy the access policy assigned by the data owner can obtain the symmetric key and recover the documents.

### 2) Unlinkability of Query

Our scheme should ensure that the trapdoors genreraterd by queries are unlinkable. That's to say, even if the data user issues two identical queries, they can be transformed into two completely different trapdoors. CSP cannot deduce the relationship between trapdoors. In the process of generating trapdoors, the data user chooses two numbers randomly and transforms the original query vector to $Q^* = (aQ, r')$. Then $Q^*$ is divided into two parts randomly according to data owner's secret key $S_{A_i}$ before encrypted by $\{M_1, M_2\}$. So even if the data user issues two identical queries, they can be transformed into two completely different trapdoors. CSP cannot deduce the relationship between trapdoors.

*3) Collusion Resistance*

In our system, we should make sure that malicious users in the system can not combine their secret key components to get more privileges they would not have otherwise.

In our scheme, the secret value of root node is covered by the component of ciphertext *CK* instead of the private key. To decrypt a document successfully, attackers must recover $Y^{r_0}$. To do this, the attacker must compute the bilinear map of $C_n$ and $S_m$. The aimed value $Y^{q_n(0)}$ can be obtained by this way. Because $u$ is a random number chosen by a data user, the secret value $Y^{r_0}$ cannot be recovered unless the user has enough private key components to satisfy the access tree which is hidden in the ciphertext. Malicious can't collude to access documents unauthorized because of the randomization of the user's secret keys.

## VI. CONCLUSION

In our work, we propose our efficient semantic and authorization search scheme ESAS to better adapt to distributed cloud storage system which achieves faster response speed comparing to existing solutions as well as enhances search capabilities. The security and efficiency analysis indicate that our system meets security requirements and has a faster search response speed. Our proposed scheme greatly improves the users' experience and is very promising and practical. In the future work, we will take dynamic updates and multi-contributor scenario into consideration.

ACKNOWLEDGMENT

This work is supported by Beijing Natural Science Foundation under grant 4182060. The corresponding author is Zhitao Guan.